\documentclass[fleqn,usenatbib]{mnras}

\usepackage{newtxtext,newtxmath} 
\usepackage{url}
\usepackage[T1]{fontenc}
\usepackage{ae,aecompl}
\usepackage{hyperref}
\usepackage{color}

\usepackage{graphicx}	
\usepackage{amsmath}	

\title[Merger remnants and their classification]{Remnants of recent mergers in nearby early-type galaxies and their classification}

\author[Giri et al. 2023]{
Gourab Giri$^{1,2}$\thanks{E-mail: gourab@iiti.ac.in},
Sudhanshu Barway$^{3}\thanks{E-mail: sudhanshu.barway@iiap.res.in}$,
Somak Raychaudhury$^{4,5,6}$\thanks{E-mail: somak@iucaa.in}
\\
$^{1}$ Department of Astronomy, Astrophysics and Space Engineering, Indian Institute of Technology Indore, Simrol 453552, India\\
$^{2}$ Department of Physics, Presidency University, Kolkata 700073, India\\
$^{3}$ Indian Institute of Astrophysics, Bengaluru 560034, India\\
$^{4}$ Department of Physics, Ashoka University, Sonipat, Haryana 131029, India\\
$^{5}$ Inter-University Centre for Astronomy and Astrophysics, Pune 411007, India\\
$^{6}$ School of Physics and Astronomy, University of Birmingham, Birmingham B15~2TT, UK\\
}

\date{Accepted XXX. Received YYY; in original form ZZZ}

\pubyear{2023}
\usepackage{cancel,soul,ulem}

\begin{document}
\label{firstpage}
\pagerange{\pageref{firstpage}--\pageref{lastpage}}
\maketitle

\begin{abstract}
We search for signatures of recent galaxy close interactions and mergers in a sample of 202 early-type galaxies in the local universe from the public SDSS Stripe82 deep images ($\mu_r \sim 28.5$  mag arcsec$^{-2}$). Using two different methods to remove galaxies' smooth and symmetric light distribution, we identify and characterize eleven distinct types of merger remnants embedded in the diffuse light of these early-type galaxies. We discuss how the morphology of merger remnants can result from different kinds of minor and major mergers, and estimate the fraction of early-type galaxies in the local universe with evidence of recent major (27\%) and minor (57\%) mergers. The merger fractions deduced are higher than in several earlier surveys. Among remnants, we find that shells are the dominant merger debris (54\%) associated with early-type galaxies, resulting from both major and minor mergers, with those characteristics of major mergers being significant (24\% of shell host galaxies). The most uncommon merger-related structures are boxy isophotes of the stellar distribution and the presence of disk fragments near the cores of galaxies. We develop a classification scheme for these fine structures that may be used to infer their likely genesis histories. The classification is primarily based on the mass ratios of the merged galaxies. This work, when combined with predictions from numerical simulations, indicates that most (if not all) early-type galaxies in the local Universe are continually evolving as a result of (minor) merger activities.
\end{abstract}




\section{Introduction}
The primary classification of galaxies into early-type and late-type ones was first introduced by \citet{Hubble1926}. Since then, significant efforts have been made to improve this classification scheme, and linking it to the evolutionary history of galaxies, by taking
more detailed structures and measurable properties into consideration  \citep{Fukugita2007, Nair2010, Cappellari2011}. Large-scale surveys, particularly those involving observations of early-type galaxies 
using integral field spectroscopy, have revealed complex structures in photometric images and in the velocity fields of stars, underlying the otherwise smooth large-scale morphology of these galaxies \citep{Zeeuw2002,Cappellari2011,Emsellem2011,Krajnovic2013,bundy2015, Krajnovic2020,Loubser2022}. These advanced observing techniques, along with the results obtained from improved image reduction and analysis tools, are now revealing a wealth of information about these intricate structures \citep{Malin1983, Jedrzejewski1987, Forbes1992, Pogge2000, Tal2009,Duc2015}. 
These morphological signatures typically reveal low surface brightness characteristics \citep{Prieur1988, Schombert1990,Cronojevic2016,Mancillas2019}, and studies suggest that they are primarily associated with mergers or close interactions between galaxies \citep{Kaviraj2010,Meusinger2017,Wang2020,Martinez-Delgado2022}. According to the $\Lambda$CDM model, the hierarchical merging of galaxies provides a central theme for galaxy evolution, and thus most galaxies, in their lifetime, have undergone several collisions which result in tidal debris \citep{Bullock2005, Cooper2010,Lotz2011,Puech2012,Wang2015,Rodriguez-Gomez2016,Hammer2018}. As the early-type galaxies attain their current form at the end of this gradual mass assembly process, one would expect a greater number of merger signatures embedded in and surrounding these galaxies \citep{Naab2006,Naab2007,Naab_IAUS2013,Pop2018,Loubser2022}. 
These remnants of close interactions and mergers may appear in the form of tidal tails, shells, broad fans, peculiar dust lanes, stellar streams \& plumes  etc., with varying spatial topologies that we will further discuss in this work. A detailed investigation of these fine structures is essential for reconstructing the history of mass assembly of a galaxy \citep[e.g.]{Nolan2007,Mahajan2009,Pop2017,Schell2017,Bilek2021a}. Numerical simulations, carried out to better understand the formation and evolution of these non-linear morphologies, also highlight the crucial link between galaxy dynamics and its evolution with the merger activities \citep{Johnston2008,Pop2018,Mancillas2019,Bilek2021}.

In groups and clusters of galaxies, the enhanced density of galaxies leads to close encounters of galaxies and often to mergers
\citep[e.g.][]{Moss2006,Tempel2017}, which is also the case in inter-cluster filaments of the cosmic web as well \citep[e.g.][]{porter2005}. The velocity dispersion of the systems plays a crucial role here, such that lower dispersions lead to higher dynamical friction in interacting systems, leading to merger. Thus, the cluster environment, and more importantly, the group environment, becomes a locus of galaxy evolution due to the higher rate of mergers and tidal interactions \citep[e.g.][]{Mulchaey1998,Mahajan2009,Clogs2018}. 
Other properties of galaxies related to the physics of close interaction and merging may also be detected, along with associated morphological signatures, in observations of these systems.
For example, mergers frequently cause higher star formation in galaxies, influencing their evolution, depending on the gas content of the merging galaxies \citep[][]{Nolan2007,gas2018,pearson2019}, which can also be associated with image features. 
Recent dedicated deep and wide galaxy surveys have made it possible to collect detailed information about these features in statistically motivated samples \citep{Martinez-Delgado2010,Duc2015,Fliri2016,Hood2018,Martinez-Delgado2021}. For our work here, we use the deep survey by Sloan Digital Sky Survey (SDSS) of the Stripe82 region \citep[The IAC Stripe82 legacy project:][]{Fliri2016}. 

Structures associated with the remnants of galaxy interactions or mergers have a certain (often characteristic) lifetime, eventually getting dissipated and buried in the resulting galaxy, generally becoming too faint to be detected. Typically, structures associated with minor mergers last a billion years, whereas structures related to major mergers have longer lifetimes \citep[see][discussed later in this paper]{Mancillas2019}. 
By searching for merger remnants using deep observations and knowing these structures' lifetimes, one can estimate the galaxy merger rate at a particular redshift (and vice-versa). There are cosmological simulations based on the $\Lambda$CDM model from which rates of galaxy mergers in the high redshift universe, and in the local universe, can be estimated \citep{Bertone2009,Hopkins2010,Rodriguez-Gomez2015}. However, the observational determination of these rates is quite challenging. The merger rate at high redshifts is usually determined from counts of interacting pairs, since gravitationally bound galaxy pairs are destined to merge within a few Gyr \citep{Patton2002,Bell2006,Patton2008,Bundy2009},
while, at low redshift, a better census of mergers can be obtained from post-merger remnant structures, which are generally of low surface brightness and lying hidden beneath the diffuse light in galaxies \citep{Heiderman2009,Casteels2013,Cronojevic2016,Mancillas2019}. 
The close pair counting process is not free from pitfalls from contamination from  projected spurious pairs, which  may introduce biases \citep[see][]{Bell2006,DePropris2007,Jian2012,Ventou2019}. The same is also true in the low redshift regime if one does not have well resolved datasets. However, while working with reasonably deep observations, counts of merger remnants lead to an efficient way of probing the merger rate and fraction of galaxies in the local universe \citep{LeFevre2000}.

This paper concentrates on finding faint merger remnants in early-type galaxies and studying their observable characteristics, in order to statistically estimate the frequency of occurrence of minor and major galaxy mergers leading to the formation of early-type galaxies in the local universe. This is important because the connection between the various parameters of an interacting system of galaxies with the observed structure of the underlying merger remnants is poorly understood \citep{Bilek2014,Kado-Fong2018,Ren2020}. The complex morphology of these remnants requires the involvement of models with several parameters to be reproduced in numerical simulations \citep[e.g.][]{Hernquist1988, Johnston2008, Hendle2015}. Thus, any correlations or characterization of merger structures found in these analyses can further provide crucial constraints on our understanding of the formation process of such peculiar embedded structures, along with finding their connection with galaxy evolution \citep[e.g.][]{Wang2020,Sola2022}.

This paper is arranged as follows: in \S\ref{data}, we describe our sample and outline the selection criteria chosen for our particular science goal. In \S\ref{iat}, we demonstrate the performance of the techniques used in this work to uncover the underlying merger-related structures. We further present the features recovered from our sample along with their observable properties in \S\ref{Merger signatures associated with early-type galaxies}. In \S\ref{Classification of merger remnants}, we propose a scheme for the classification of these merger structures, indicative of their possible merger origin.  We estimate the merger fraction of galaxies in the local universe in \S\ref{Finding merger fraction of ETGs in the local universe}. We then compare our work with the predictions from theoretical studies in \S\ref{Synergy with theoretical predictions}, and list the key conclusions obtained in \S\ref{Conclusions}.

\section{Sample  selection and data} \label{data}
 
In this work, we search for low surface brightness features in a sample of early-type galaxies. For this purpose, we make use of the deep imaging in the SDSS Stripe82 (S82 hereafter) from the {\it IAC Stripe82 legacy project}  \citep{Fliri2016}. The S82 is a region of the sky delineated by $-50^{\circ} \leq {\rm R.A.} \leq +60^{\circ}$ and $-1.25^{\circ}\leq {\rm Dec.} \leq+1.25^{\circ}$, covering $275~degree^{2}$ along the celestial equator in the direction of the southern galactic cap. The IAC Stripe82 legacy  project provides deep images, co-adding multi-epoch observations to reach very faint surface brightness limits (3$\sigma$ limits of 28.5 per arcsec$^{2}$) in the $r$-band. 
To obtain morphological information for the galaxies found in the S82 deep images, we have used the {\it Galaxy Zoo 2 Strip82 co-added depth catalogues} \citep{Willett2013}, which provide visual morphology for $\sim$39,530 galaxies and classify them into elliptical and spiral morphological types. We then select only the galaxies classified as elliptical (amounts to $\sim$72\% of the catalogued galaxies). The catalogue further divides elliptical galaxies (E) into three classes: E$_r$, E$_i$ and E$_c$, where the subclasses $r$,  $i$,  and $c$  refer to round, intermediate, and cigar-shape, respectively. Since from this crude classification, it is not possible to distinguish between Es and S0s, we will refer to our galaxies as early-type or ETG hereafter.

To obtain parameters such as position, Petrosian radius and redshift of these galaxies, we used the table {\it SDSS metadata for GZ2}. We applied a cut-off in redshift $z\leq 0.05$ to retain only the nearby early-types. In order to search for low brightness features, we select galaxies with large enough angular extent by putting another cut in Petrosian radius, i.e. $R_{p90}\geq 10\arcsec$. This gives us a sample of 566 ETGs, 
which we use for this study.

We have also used a sample of 55 bright elliptical galaxies taken from the OBEY survey (Observations of Bright Ellipticals at Yale)  as a control sample \citep{Tal2009}. We use this sample to check the compatibility of the techniques used in this study. 

\section{Image analysis techniques} \label{iat}

To uncover the low surface brightness features that are either buried in the smooth and often symmetric dominant luminosity distribution of the galaxies or in the sky background of deep optical images from SDSS S82, we use two methods elsewhere proven useful in identifying various low surface brightness features. They are:
(a) {\it unsharp masking} and (b) {\it model subtraction}. We use these two different techniques in our images to help us identify low surface brightness signatures by comparing the residual images and to identify possible artefacts that may be the outcome of the techniques themselves. Both these techniques have their advantages and shortcomings, which we discuss below. 

\begin{figure}
\centering
\includegraphics[scale=0.68]{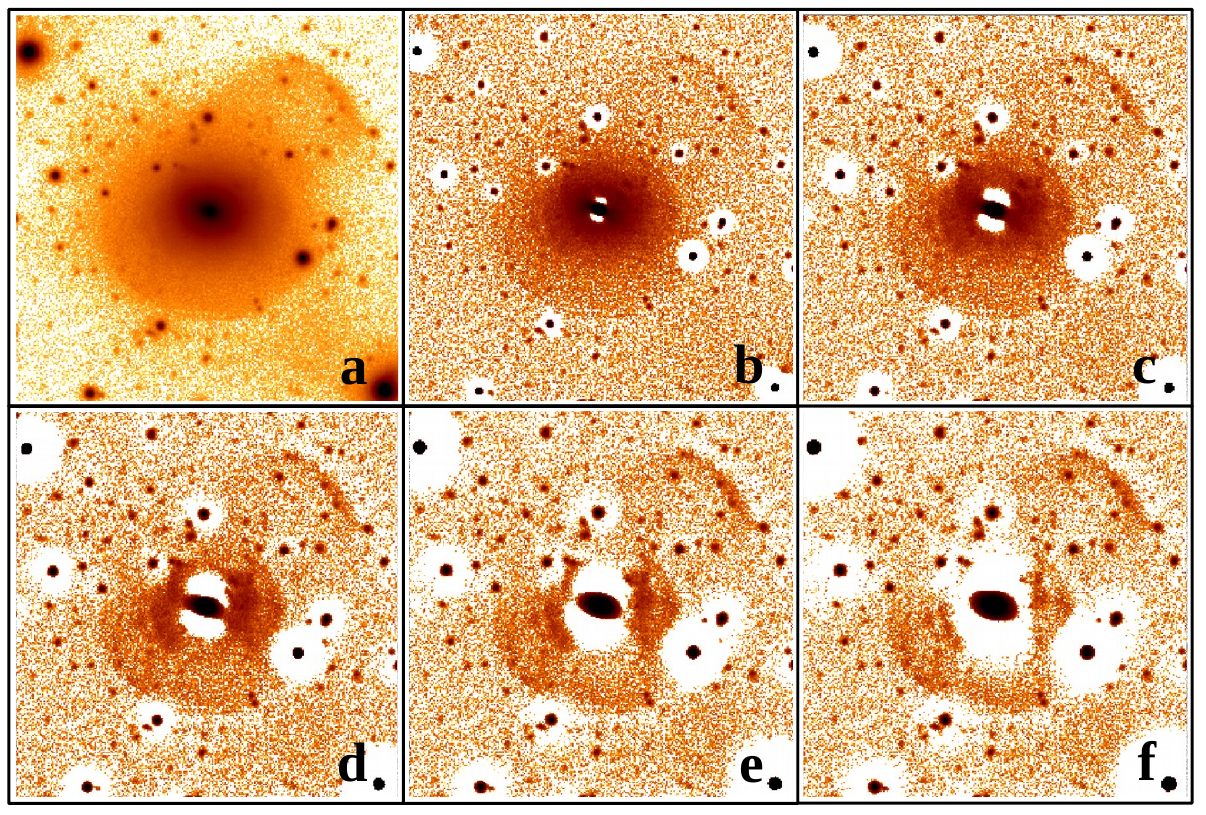}
\caption{Unsharp masking technique with different choices of Gaussian kernel ($\sigma$): (a) original galaxy image, (b)--(f) unsharp mask output images with  $\sigma = 3$,  $\sigma = 5$, $\sigma = 7$, $\sigma = 9$ and $\sigma = 11$, respectively. The contrast enhancement parameter $ C = 5$ for all the cases. It can be noticed from the above images that a higher value of $\sigma$ can distort the features near the centre of the galaxy. In contrast, a lower value of $\sigma$ is unable to detect the outer structures sharply. The tiny blobs in the images are the over-subtracted foreground or background luminous sources.}
\label{fig:gauss}
\end{figure}

\subsection{Unsharp masking}
In this technique, we first make a smooth image of the original galaxy using a Gaussian kernel of size $\sigma$ using  Image Reduction and Analysis Facility (IRAF)\citep{Tody1986}.  Then increasing the intensity of the original galaxy image and also of the created smooth image by a contrast enhancement parameter $C$, we have subtracted them according to \citep[e.g.][]{Pogge2000,Peng2002}
\begin{equation}
E_I=C*G-(C-1)*S,
\end{equation}
where $E_I=$ the Enhanced image, $C$ = Contrast factor ($C>1$), $G$ = the original Galaxy image and $S$ = the Smoothed image (notes on Contrast Enhancement using Digital unsharp masking contents are available online \footnote{Contrast Enhancement: \url{http://www.astronomy.ohio-state.edu/~pogge/Ast350/Unsharp/}}).
This way, the two-dimensional galaxy profile, smoothed with a Gaussian kernel, is removed to make the low surface brightness features more prominent in the enhanced image. The values for Gaussian kernel size ($\sigma$) and contrast factor ($C$) must be manually chosen to enhance the contrast.

In this regard, Fig.~\ref{fig:gauss} shows the residual images obtained after performing the above process on one of our images with different values of kernel width, and for a fixed contrast parameter ($C = 5$). These images show that even though higher values of sigma will enhance the sharpness of the extended features,  it leads to over-subtraction near the centre of the galaxy, eventually distorting the central structures. On the other hand, a low value of sigma enhances small features, making larger features less prominent. These comparisons have also been made by using different values of $C$, and we find that changing the value of $C$ does not affect the enhanced image as much as $\sigma$ does. 

We thus construct images using this unsharp mask technique for each deep S82 co-added image of our sample of 566 galaxies, using a range of kernel sizes ($\sigma$). The residuals from these unsharp-masked images reveal significant embedded spiral structures in 221 galaxy images, which leads us to believe that these spiral galaxies were misclassified as `ellipticals' in Galaxy Zoo~2, where only the original intensity images were used for morphological classification.  

\begin{figure}
\centering
\includegraphics[scale=1.0]{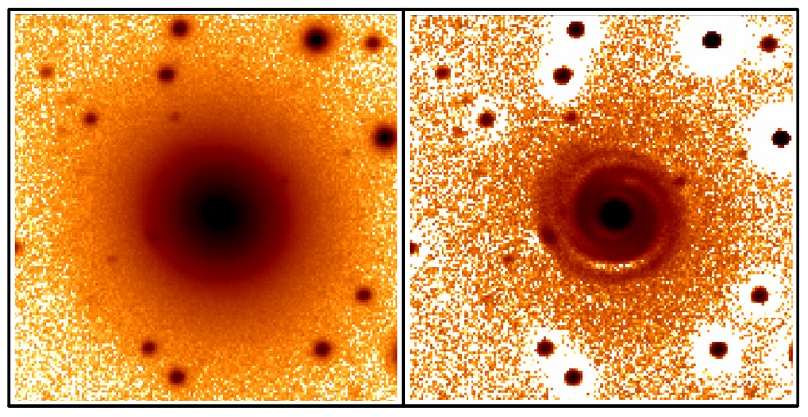}
\caption{An example of the spiral structure (right) was revealed after applying the unsharp masking technique on the galaxy. 
SDSS J211603.77+010106.0 (left) was identified as an elliptical galaxy in the Galaxy Zoo 2 data.}
\label{fig:spiral}
\end{figure}

Fig.~\ref{fig:spiral}  shows an example of such misclassification, where a galaxy was classified as an `elliptical' in Galaxy Zoo 2, but the residual image from unsharp masking shows distinctive spiral features. In other images,  we find irregular galaxies and edge-on-disk galaxies that are misclassified as ETGs. We exclude such galaxies from our subsequent analysis, reducing our sample to 202 galaxies classified as elliptical by Galaxy Zoo~2 and showing no sign of large-scale spiral features in the residual images.

\begin{figure}
\centering
\includegraphics[scale=1.0]{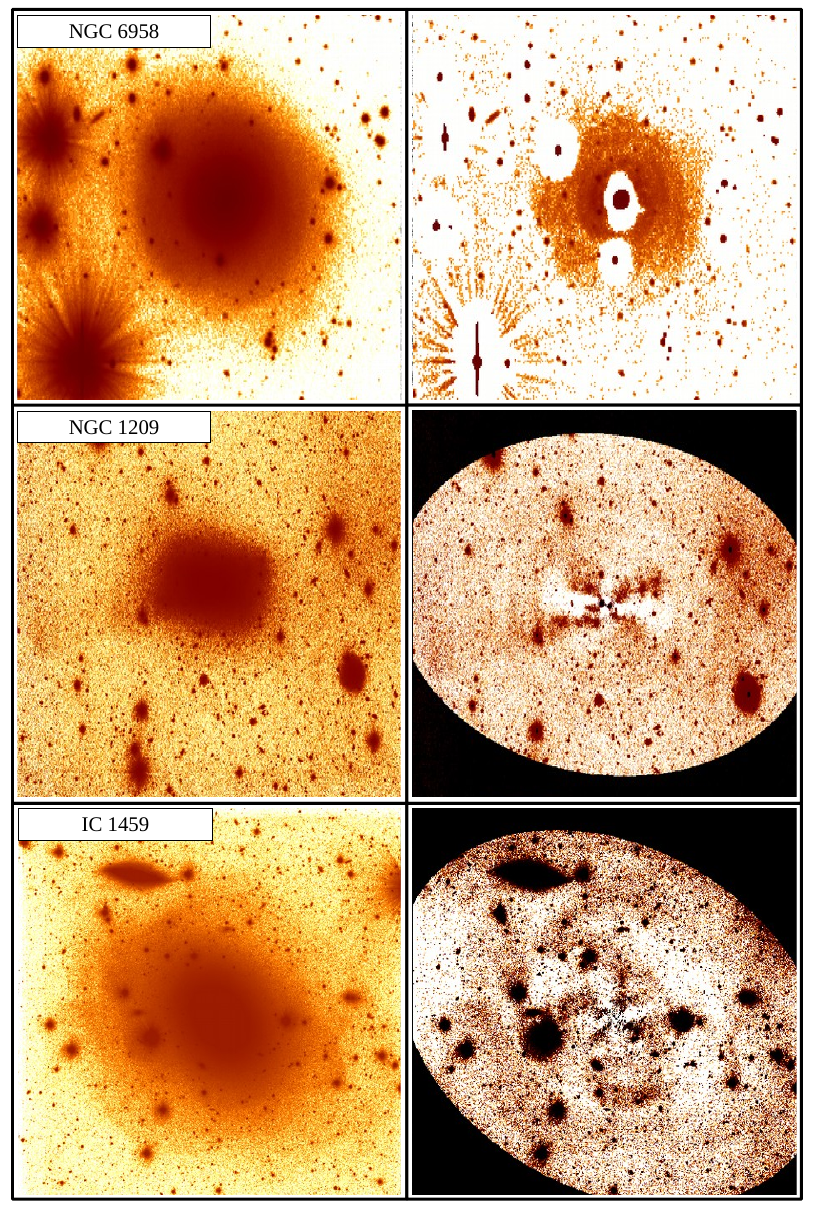}
\caption{Image analysis techniques that we used to recover the structures from the original images in this work. The galaxies shown here are from \citet{Tal2009}. The left column represents the original galaxy images, and the right column represents the residual images. The retrieved structures for three galaxies, i.e. NGC~6958, NGC~1209 and IC~1459, are shown, where inner shells (top right), X-structure with extended arm  (middle right), and inner and outer shells (bottom right) are recovered using unsharp masking (top) and model subtraction technique (bottom two).}
\label{fig:obey}
\end{figure}

\subsection{Model subtraction}\label{Model subtraction} 
 
In order to model the galaxy light distribution to $r$-band images,
we created model galaxy images with elliptical isophotes using the task ELLIPSE in the STSDAS package available in IRAF \citep[see, e.g.][]{Barway2005}. The output from the ELLIPSE task was fed into the task BMODEL to obtain a smooth model of the galaxy, which was subtracted from the original image to get a residual image.
Noise features, such as stars, bad pixels, etc., were identified in the first round of the fitting procedure and masked in the next run.

The input parameters, such as position angle, ellipticity, and the centre of the ellipse, were allowed to vary during the fitting. We repeated the fitting procedure by fixing or varying one or all of these parameters (i.e. position angle, ellipticity, and ellipse centre) to check the stability of fit and how well the features are detected in residual images. In addition, we applied this technique to the control sample of 55 bright elliptical galaxies from the OBEY survey. In Fig.~\ref{fig:obey}, we show the corresponding output for three galaxies, NGC~6958, NGC~1209 and IC~1459, demonstrating that we can recover the features reported for these galaxies by \citet{Tal2009} using our adopted techniques. 

We found that in order to extract the asymmetric features that we report in this work, we require allowing the parameters position angle, ellipticity, and the centre of the ellipse to vary while fitting the models.
\\
\\
\noindent In summary, we note that two methods have proven useful to unravel the low surface brightness features embedded in the smooth light distribution of the galaxy or the sky background of deep images: \textit{unsharp masking} and \textit{model subtraction}. We first used the unsharp masking technique in all our images (i.e., 566 Es (as classified by Galaxy Zoo 2), obtained from SDSS S82). We found that several galaxies were misclassified as ellipticals by Galaxy Zoo 2. As a result of excluding such galaxies from our sample, the number of galaxies of ETG morphologies has been reduced to 202, which is the final sample used further in this study. We then apply the model subtraction technique in these 202 galaxies to further detect the merger signatures by comparing the results with unsharp masking detection.

\section{Merger signatures associated with early-type galaxies} \label{Merger signatures associated with early-type galaxies}

In this work, we look for signatures of recent mergers in our early-type galaxy sample using the techniques mentioned above. We attempt to identify these features and propose a classification system indicative of possible origin histories. 

\subsection{Structural Characteristics} \label{structural characteristics}
\begin{figure*}
\centering
	\includegraphics[scale=0.86]{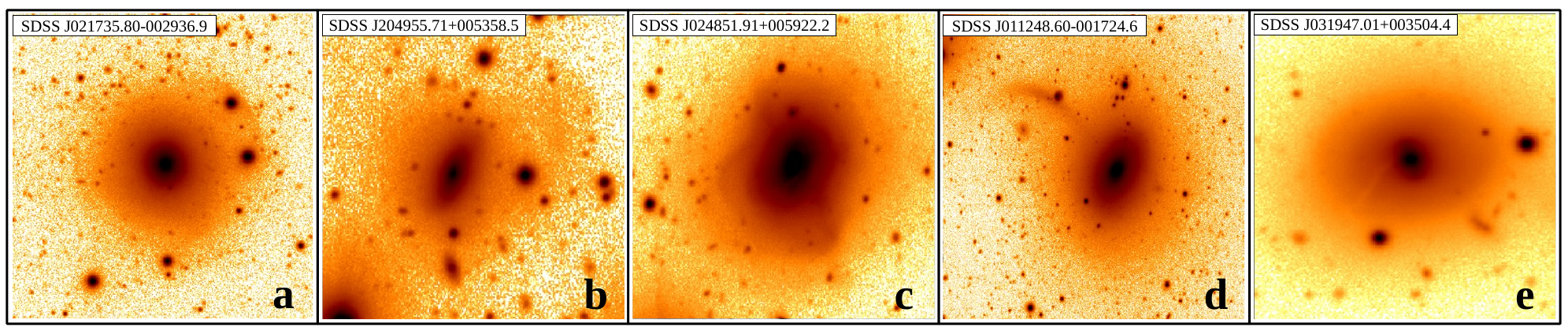}
    \caption{The outer structures in the galaxy images were obtained by adjusting only the intensity scale and are identified as: outer shells (a), tidal fan (b), stellar streams (c), tidal tail (d) and dust lane (e), respectively.}
    \label{fig:seen_from_original}
\end{figure*}

We have identified merger structures by visual inspection, given that automated processes can have issues which raise misleading results \citep[e.g.][]{Duc2015,Jackson2021,Martinez-Delgado2021,Sola2022,Lagos2022}. A subset of our sample is also cross-checked by individual co-authors, with a high level of agreement, in determining the morphology of the structures. We recover a few prominent outer merger remnants by only adjusting the intensity scale of the galaxy, as has been demonstrated in Fig. \ref{fig:seen_from_original}.
The structures obtained using the image analysis techniques are highlighted in Fig.~\ref{fig:features_seen_in_technique}.

\subsubsection{Tidal tails}
Tidal tails result from the merger of a disk galaxy with a companion  \citep[e.g.][]{Schombert1990}. They can result from major or minor mergers, with subtle differences in the outcome. Tails created from major mergers in disk-disk galaxy collisions become broad, diffuse, long and curved. In many cases, simulations found the presence of at least two tails in the post-merger stage \citep[e.g.][]{Toomre1972,Hibbard1995}. In contrast, faint, short, thin tails are thought to be the outcome of minor mergers \citep[e.g.][]{Feldmann2008}. The detailed processes involved in forming these structures are not fully understood \citep{Ren2020}.

In our sample from S82, we found only nine galaxies showing tidal tail features (5\%),
which is less than the value quoted in \citet{Tal2009}, who analysed a similar sample. It is expected, as they did not explicitly consider stellar streams and plumes (discussed later) as a separate merger feature. Seven of the ETGs, show broad, diffuse and curvy tails, indicating a recent major merger event (Fig.~\ref{fig:major_minor_difference1}). These prominent features can also be detected in the original images, as shown in Fig.~\ref{fig:seen_from_original}.
\subsubsection{Tidal fans}
\begin{figure}
\centering
	\includegraphics[scale=1.0]{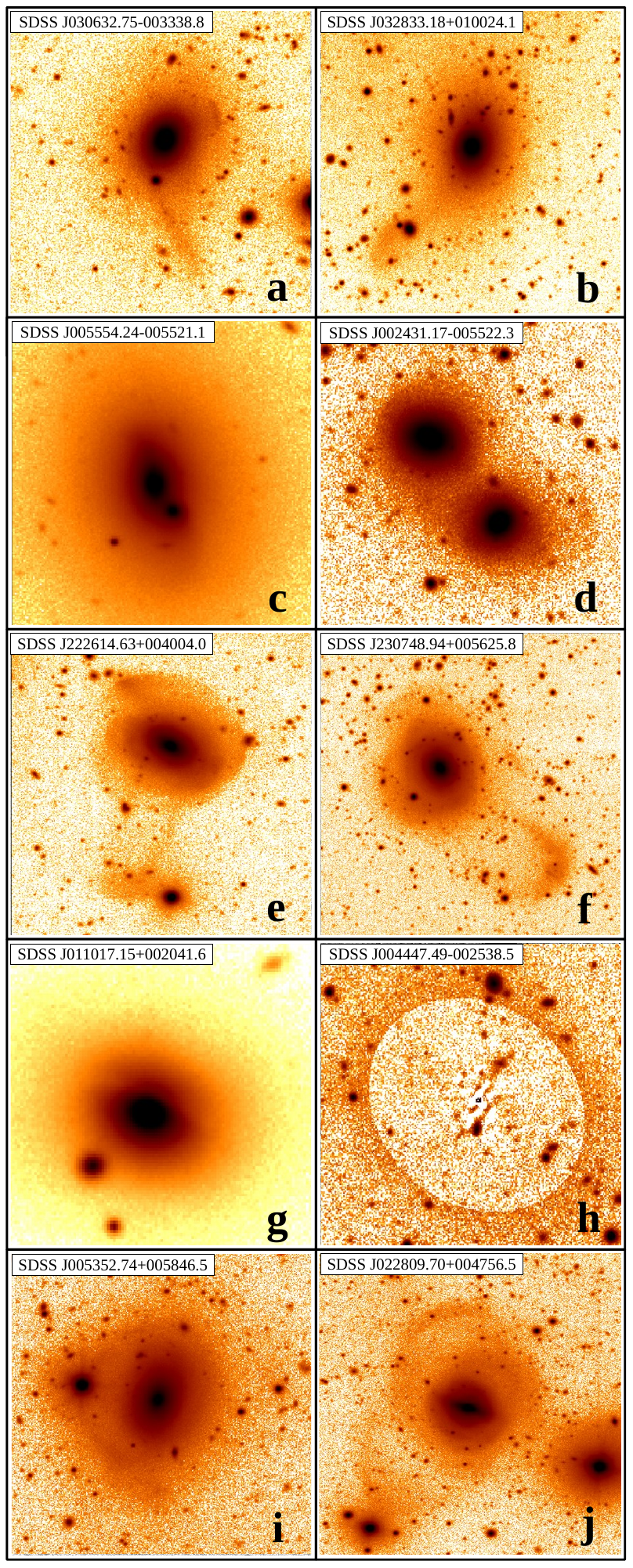}
    \caption{The morphology of merger structures originating from the minor (left column) and the major (right column) merger, showing subtle differences (see Section \ref{structural characteristics}). The structures represented here (from top to bottom) are tidal tails (a: minor merger, b: major merger), tidal fans (c: minor merger, d: major merger), shells (e: minor merger, f: major merger), dust lanes (g: minor merger, h: major merger) and stellar streams and plumes (i: minor merger, j: major merger).}
    \label{fig:major_minor_difference1}
\end{figure}

Similar tail-like features occur during the close encounters of galaxies, in particular in merger sequences of pairs of early-type galaxies. Since stellar velocity dispersion in early-type galaxies is higher than in late-type systems, these features end up broadened, and appear to have fan-like structures, often known as tidal fans \citep{Colina1995}. As examples, we can cite the well-known galaxy pairs NGC750/NGC751 and NGC7236/NGC7237. These features have also been produced in simulations involving dry mergers \citep[e.g.][]{Combes1995}, and are seen in both minor and major mergers. Interestingly, in minor interactions, the lower-mass galaxy is expected to show the more prominent fan feature (see Fig.~\ref{fig:major_minor_difference1}). The stellar distribution having elongated lumpy morphologies, sometimes found in the outskirts of recently merged galaxies, is also a signature of tidal fans as seen in \citet{Greco2017} \citep[see][for their plausible accretion origin]{Bullock2005}.

We found 18 galaxies (nearly 9\%) in our sample, showing this broad fan feature. Several of these structures are apparent in the original images as well, even before unsharp mask or model subtraction is applied (Fig.~\ref{fig:seen_from_original}, \ref{fig:major_minor_difference1}(e)). 
\subsubsection{Shells}
Shell structures, found in early-type galaxies, are concentric, interleaved and have sharp-edged features (see Fig.~\ref{fig:seen_from_original}) consisting mainly of stars \citep{Pence1986} and sometimes gas \citep{Schiminovich1994,Mancillas2019a}. These features occur in a wide range of galactocentric radius, and are mostly found in ETGs in low-density environments \citep{Malin1983a,Malin1983,Bilek2014,Bilek2021} 

Since their discovery in early unsharp-masked deep sky survey images \citep{Malin1983}, shells of various kinds have been widely explored in theoretical studies and are believed to have their origin in mergers.
\citep[e.g.][]{Dupraz1986, Hernquist1988, Turnbull1999, Pop2017, Mancillas2019}. 
The merger scenarios generally feature accretion events between a primary and a less massive companion galaxy. The stellar orbits that eventually are associated with the shells are stretched into thin sheets in phase space, wrapped around an invariant gravitational potential surface \citep{Quinn1984, Fort1986, Nulsen1989, Turnbull1999, Ebrova2012}. 
However, there are equally compelling scenarios featuring major mergers that can lead to shells \citep{Schweizer1982, Hernquist1992, Janowiecki2010, Mancillas2019}. These are seen in both analytical and numerical exercises, and it is becoming clear that shells resulting from major mergers are not unusual \citep[e.g.][]{Pop2018,Kado-Fong2018}.
In this paper, we have considered both the major and minor merger origins for shell formation. In cases where it is clear that the large-scale features of the merger point to a major merger, we have classified them as such (Fig~\ref{fig:major_minor_difference1}). 

We found shell-like structures in 109 early-type galaxies, which amounts to 54\% of the sample. It falls between the incidence quotes in the samples of \citet{Schweizer1985} (i.e. 44\%) and that of \citet{Rampazzo2020} (i.e. 60\%). However, we claim that our result is statistically more significant, as we have 202 ETGs in our sample in total, compared to the 36 Es and 20 ETGs, respectively, in the above studies. 
One should also note that this value is noticeably higher, considering several recent studies by \citet{Tal2009,Duc2015,Pop2018}. Among these 109 ETGs where shells have been found, 26 of them feature shell structures that have a major merger origin. It is equivalent to 13\% of the entire sample (i.e. within 202 ETGs) and 24\% of the shell host galaxies (i.e. within 109 ETGs hosting shells). This number is slightly higher than the number reported by \citet{Kado-Fong2018} (i.e. $15\%\ \pm\ 4.4\%$ of shell host galaxies) who labelled merger origins to shell structures based on their ($g-i$) colours. Their lower detection can be attributed to several factors, including cosmological surface brightness dimming, lower special resolution for high redshift galaxies, and issues in background subtraction for some low redshift galaxies, as highlighted by the authors.
Both unsharp masking and model subtraction techniques have proven to be helpful in finding the inner and outer shells of galaxies (Fig.~\ref{fig:features_seen_in_technique}). The difference is the fraction of identified low-surface brightness features, such as shells, can be attributed to these methods.
\subsubsection{Collisional rings}
Close interaction and mergers of galaxies can also result in ring-like features, often referred to as collisional rings \citep{Theys1976,Few1986,Barway2020}.
They are not the same as resonant rings, which are formed due to the dynamical resonances 
within the disks of galaxies, raising morphologies like nuclear rings, inner and outer rings (typically enclosing a bar or a spiral arm) or pseudo rings (compact spiral arms)  \citep[e.g., see][]{Buta2017a, Buta2017}. The resonant rings are most commonly associated with the disk galaxies \citep{Buta1996}, which lie outside the topic of our interest.

A minor merger between a disk galaxy and a small companion galaxy is thought to be the major cause of collisional ring features, subject to a combination of conditions that lead to this morphology \citep[e.g.][]{Madore2009}. Numerical simulations showed that a high-speed head-on collision is required between the disk galaxy and the less massive companion to form this structure, with the companion traversing close to the centre of the disk \citep{Theys1977,Bekki1998,Renaud2018}. 
As a result of this, as well as due to their shorter lifetime of existence, the incidence of such features is quite low \citep{Buta2015,Madore2009,Renaud2018}.
The collisional rings can have morphologies like polar rings \citep{Reshetnikov1997, Bekki1998}, Cartwheel-type rings \citep{Fosbury1977}, rings with an off-centre nucleus \citep{Lynds1976}, empty rings, or rings with dominant knots \citep{Theys1976, Theys1977, Madore2009}. 

In our study, we find that 5\% of galaxies in our sample show collisional ring structures, and unsharp masking is the better technique for finding these features (Fig.~\ref{fig:features_seen_in_technique}). Among these nine galaxies having collisional ring morphology, one shows a polar ring-type feature, one a cartwheel-type ring feature, two show off-centred rings, and the rest are deformed rings (Fig.~\ref{fig:major_minor_single_one}).
\subsubsection{X-shaped structures} 
Among the various complex long-lived features produced by minor mergers,  X-shaped structures have been studied in many systems \citep[e.g.][]{Mihos1995, Tal2009}.
A 2D elliptical model of light distribution subtracted from galaxy images brings out these structures  (see Fig.~\ref{fig:features_seen_in_technique}). The formation of these structures in S0 and early-type spiral galaxies follows the accretion of a small satellite galaxy, followed by the buckling instability of bars \citep{Whitmore1988,Portail2015}. The formation mechanism of X-structures in elliptical galaxies is more complex, but nevertheless involves a minor merger \citep[see the numerical work by][]{Hernquist1989}. We note that this structure differs from the X-shaped morphology seen in active extended radio galaxies, which involves a complex radio jet evolution \citep{Giri2022B,Giri2022A}.

In our sample, 7\% of the galaxies show this feature, one of which (SDSS J024144.38+010544.7) is shown in Fig.~\ref{fig:major_minor_single_one}.
\subsubsection{Stellar streams and plumes}
Long-term remnants that persist after mergers have occurred include deformed arcs, luminous plumes, stellar flows and loops, filamentary structures and similar features \citep{Bullock2005, Martinez-Delgado2010, Janowiecki2010, Paudel2013,Duc2015, Kado-Fong2018}. Here we group them as stellar streams and plumes. These are typically low surface brightness features and need deep imaging to be detected \citep[see the numerical work of][]{Mancillas2019}. These non-linear features have no particular spatial topologies \citep{Martinez-Delgado2021}, which distinguish them from the tidal tail structures \citep{Ren2020}. Furthermore, \citet{Sola2022} have pointed out that stellar streams are generally thinner and less intense than tidal tail features, as they primarily originate from minor mergers \citep[see also][]{Duc2015, Mancillas2019}.

In our work, we find stellar streams and plumes as the second-most frequently occurring after-merger feature (17\%). The more prominent of these stellar streams can be easily detected in the original images without any special processing, as in the example shown in Fig.~\ref{fig:seen_from_original}. These structures are sometimes associated with major mergers. An example is shown in Fig.~\ref{fig:major_minor_difference1}, where we detected a central disk-like stellar distribution indicative of a recent major merger (discussed next). The return of the tidal material to the disk after a major merger event might produce such structures, as found by \citet{Hibbard1995}.
\subsubsection{Boxy and disky isophotes}
The stellar distribution of ETGs occasionally reveals boxy or disky isophotes, typical consequences of major mergers. It is believed that the most massive ellipticals are formed from the major mergers of gas-poor galaxy collisions, often known as dry mergers. These sometimes lead them to have boxy isophotes \citep{Heyl1994, Naab2006, Graham2012, Goullaud2018}. 
On the other hand, low mass ellipticals and S0s, which are thought to be the products of major mergers of gas-rich disks, can result in remnants having disky morphologies with considerable angular momentum \citep{Scorza1995, Lauer1995,Kaviraj2010,Emsellem2011}. Often, these lead to disky remnants embedded near the centre of the galaxy, resulting in the overall diskiness of the isophotes \citep[e.g., the simulation of][showcasing such disk formation]{Lotz2008}. These disks can often be uncovered by subtracting a model of the smoothed light distribution, as discussed in Section \ref{Model subtraction}. 

Although we found only one galaxy showing boxy isophotes in our sample, the number of galaxies showing disky isophotes is 28 (see, for example, Fig.~\ref{fig:major_minor_single_one}). In Fig.~\ref{fig:major_minor_difference1}(j), the embedded stellar disk is revealed in the original image. The presence of such disks inside ETGs has also been found in several other studies \citep[e.g.][]{Buta2015}. The rarity of elliptical galaxies with boxy isophotes \citep[in line with the result of][]{Tal2009} indicates that major mergers between gas-poor galaxies are less frequent in the local universe.

\begin{figure*}
\centering
	\includegraphics[scale=0.9]{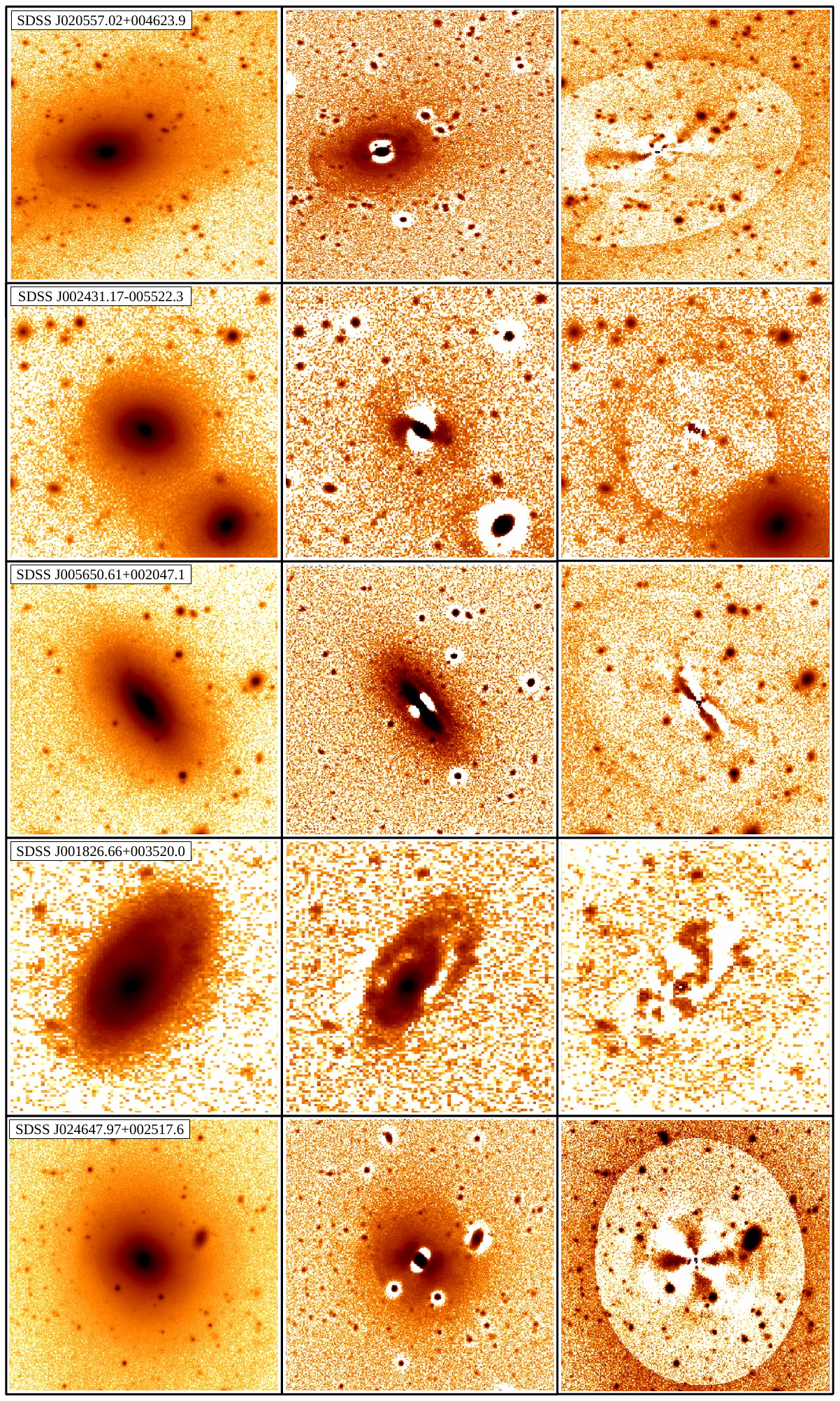}
	\caption{Left column: original $r$-band galaxy images showing relatively smooth light distribution. Middle column: unsharp masked images revealing several merger structures hiding inside the galaxy's smooth and dominant stellar distribution, as shown on the left. Right column: residual images of model subtraction technique revealing hidden asymmetric stellar distributions. The merger structures showcased here are (from top to bottom) multiple shells, embedded disk fragments, dust disk, collisional ring, and X-structure with shells, respectively. By comparing the images, it can be indicated that individual techniques have advantages and shortcomings in recovering the merger structures with varying spatial topologies (see Section \ref{structural characteristics} for details).}
    \label{fig:features_seen_in_technique}
\end{figure*}
\subsubsection{Dust disk}
The presence of prominent dust lanes in early-type galaxies (see Fig.~\ref{fig:seen_from_original}, \ref{fig:features_seen_in_technique}) is thought to result from the collision of gas-rich disks \citep{Van2015}. In these cases, the distribution of dust can be found from the outskirts of a galaxy to the very centre \citep{Raimundo2021}. Both minor and major mergers involving gas-rich disk galaxies can produce these features. However, dust disks created from major mergers are found near the core of the resultant galaxy, since these are typically observed after a long time after the merger event \citep[e.g.][]{Tran2001, Lauer2005,Ji2014,Duc2015,Wang2020}. 
On the other hand, outer dust disks are found primarily in minor merger or accretion events \citep{Graham1979, Struve2010, Neff2015}. An example of this difference is shown in Fig.~\ref{fig:major_minor_difference1}, where, in order to find the central dust lane, we had to remove the smooth light distribution of the host galaxy (e.g. SDSS J004447.49-002538.5), compared to the outer dust lanes (e.g. SDSS J011017.15+002041.6), which are evident in the original galaxy images (see also Fig. \ref{fig:seen_from_original}).

We found only 22 galaxies (11\%) showing the presence of dust lanes, of which 13 possess signs of central dust lanes. This number is fewer than that found in some previous merger studies \citep{Tran2001, Goullaud2018}, where deep and high-resolution HST images have been analysed, which might indicate that in our sample, the images are not deep enough to uncover such dust features.
\subsubsection{Disk fragments near the core}
The presence of a disk, or a disk fragment, often showing evidence of spiral features, embedded near the core of an early-type galaxy is considered a rare merger signature \citep{Lauer1995}. A possible scenario for its origin is that of a gas-rich disk falling into an early-type galaxy in a recent major merger event \citep{Li2018}.

We observed the presence of similar structures in only three galaxies in our sample, in the form of embedded spiral disks near the core of a dominant smooth elliptical light distribution. A prominent example of this is shown in Fig.~\ref{fig:features_seen_in_technique}. These are better revealed by the unsharp masking technique (e.g. SDSS J002431.17-005522.3). Another example is shown in Fig.~\ref{fig:major_minor_single_one}, where other merger structures, e.g. sharp outer shells, were also detected in addition to a disk feature. The peculiar composition and morphology of such sources warrant further investigation.
\subsubsection{Isophote twisting}
\begin{figure*}
\centering
	\includegraphics[scale=1.15]{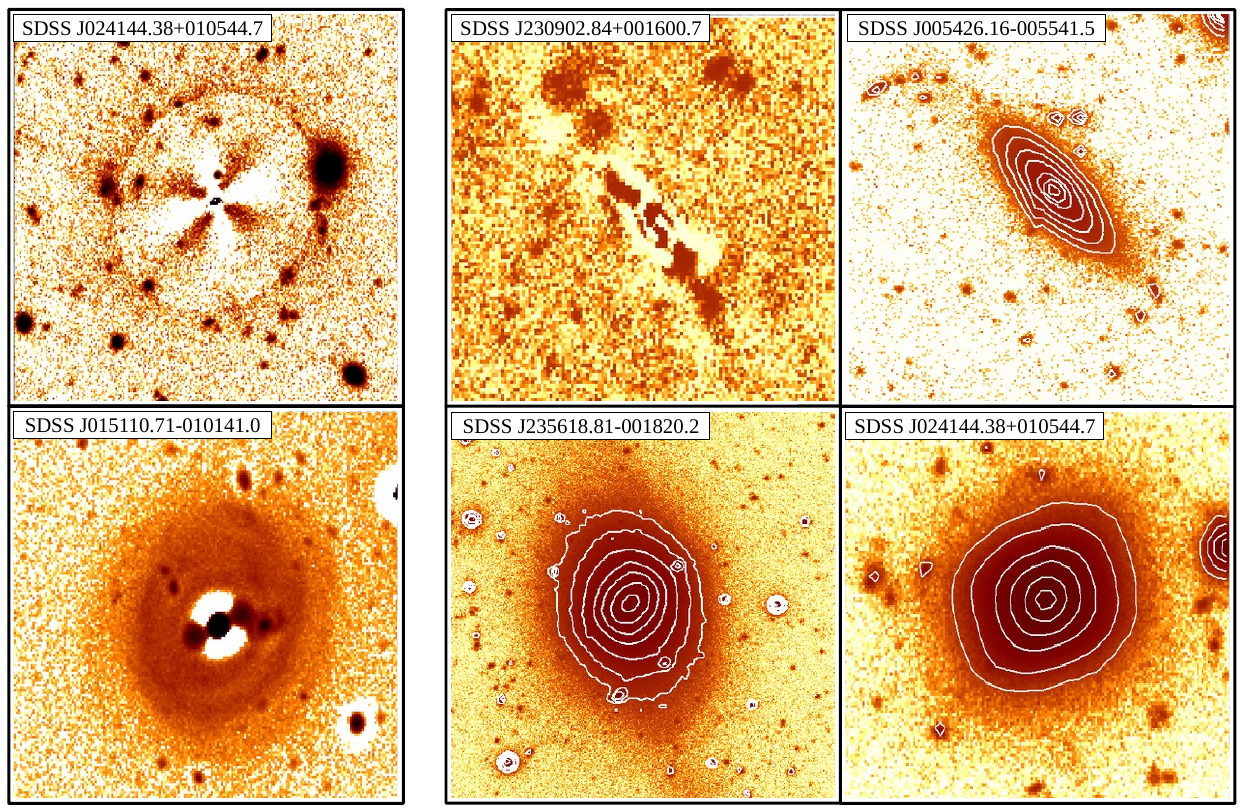}
    \caption{The merger remnants with a minor merger origin (left column) and a major merger origin (right column). The left column show X-structure (top; model subtracted image) and collisional ring (bottom; unsharp masked image) structure. The right column shows the features: disk fragment near the core (top left; model subtracted image), disky isophote (top right), isophote twisting (bottom left) and boxy isophote (bottom right), respectively. See Section \ref{structural characteristics} for more details.}
    \label{fig:major_minor_single_one}
\end{figure*}

In some ETGs, an isophotal analysis of their light distribution reveals a systematic change in the position angles of the principal axes of the isophotes, known as twisting  \citep{Gerhard1983}.
An example is given in  Fig.~\ref{fig:major_minor_single_one} (SDSS J235618.81-001820.2).
It has been proposed that such features are products of major mergers \citep{Ebrova2021}. However, there is disagreement over their process of formation \citep{Lagos2022}. \citet{Loubser2022} have recently shown that the high-mass ellipticals typically are slow rotators, contrary to low-mass ETGs, as they primarily form from a series of dry mergers.  In our sample, we found only 12 galaxies (6\%) showing this feature, slightly lower than fractions quoted by others \citep{Sanders2015, Goullaud2018}, but similar to a few others \citep[e.g. 9\% among ETGs in][]{Krajnovic2011}. 

These 12 galaxies (found in our S82 sample) show the presence of various other merger signatures, along with the isophotal twisting of the galaxy's light distribution \citep[similar to][]{Ebrova2021}. However, we cannot conclusively state that the same merger has generated all these detected features of such ETGs. If the same merger originated the features observed, then we should expect more twisted isophotes in ETGs than what has been obtained here. It asks for further studies in order to better understand the constraints involved in the production of such a twisting state of the isophotes of these galaxies \citep[e.g.,][]{Loubser2022,Lagos2022}. 

Out of these 12 samples, we obtained a majority of ETGs ($\sim 8$) showing the presence of multiple shells (> 3) associated with the twisting isophotal characteristics. It is contrary to the suggestion of \citet{Ebrova2021}, who find an overabundance of such systems. It further indicates that these galaxies could be slow rotators, which are likely to host multiple shells than fast rotators \citep[see,][]{Valenzuela2022}.
\subsubsection{Disturbed galaxies}
\begin{figure}
\centering
	\includegraphics[scale=1.0]{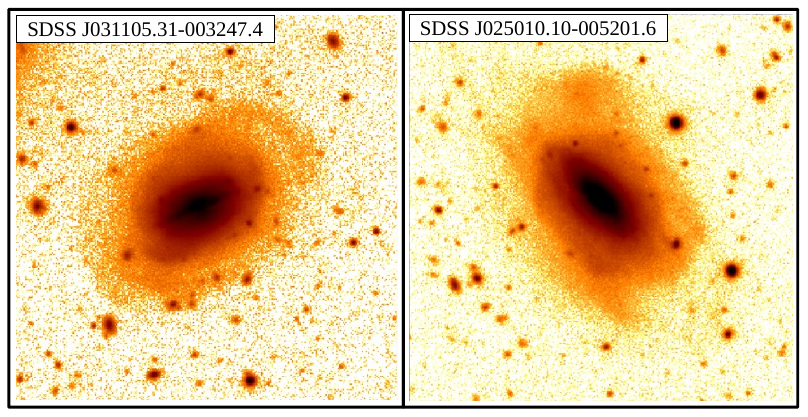}
    \caption{Examples of Early-type galaxies (original $r$-band images) with highly asymmetric stellar distributions, indicating a recent merger.}
    \label{fig:disturbed_ETGs}
\end{figure}

Signatures of major mergers can often be very complex, and post-merger remnants can be left in a highly disturbed state \citep{Ji2014,Wang2020}, far from any symmetric configuration (see Fig. \ref{fig:disturbed_ETGs}). It can also occur in minor mergers where the components have comparable mass \citep{Mancillas2019}. In this study, we found 18 galaxies with apparent asymmetrically disturbed stellar distribution, with 11 (6\%) being in a highly disturbed state indicative of a recent major merger event \citep{Tal2009}. We note that some of these ETGs possess complex central structures with a very complex spatial topology near the core, a criterion that we also used to categorize the galaxies as disturbed ones. 

\subsection{Frequency of incidence of various features} \label{Statistics of merger structures}

\begin{table}
	\centering
	\caption{The various merger features identified in the sample of ETGs from S82. The fractions reported in Column 4 show how many of a particular feature originate from minor or major mergers. So, a 100\% fraction means it has only minor or major merger origin.}
	\label{tab:Major_minor_structure_availability}
	\begin{tabular}{|l|c|c|c|} 
		\hline
		\textbf{Structures} & \textbf{Merger} & \textbf{No. of} & \textbf{Fraction} \\
		 & \textbf{type} & \textbf{ETGs} &  \\
		\hline
		\hline
        Tidal tails & minor & 2 & 22\% \\
         & major & 7 & 78\% \\
        \hline
         Tidal fans & minor & 17 & 94\% \\
         & major & 1 & 6\% \\
        \hline
        Shells & minor & 83 & 76\% \\
         & major & 26 & 24\% \\
        \hline
        Collisional rings & minor & 9 & 100\% \\
        \hline
        X-structures & minor & 13 & 100\% \\
        \hline
		 Stellar streams  & minor & 22 & 65\% \\
         \& plumes & major & 12 & 35\% \\
        \hline
        Boxy isophotes & major & 1 & 3\% \\
        Disky isophotes & major & 28 &97\%\\
        \hline
         Dust disks & minor & 9 & 41\% \\
          & major & 13 & 59\% \\
        \hline
         Disk fragments & major & 3 & 100\% \\
        \hline
        Isophote twisting & major & 12 & 100\% \\
        \hline
       Disturbed & minor & 7 & 39\% \\
       galaxies & major & 11 & 61\% \\
       \hline
	\end{tabular}
	\\
    \raggedright Note: Column (1)  type of structure, Column (2)  merger type, Column (3) number of early-type galaxies, Column (4) fraction in each merger type. 
\end{table}

\begin{table}
	\centering
	\caption{The distribution of ETGs in our sample according to merger type. The percentage (Column 3) shows the fraction of galaxies falling into these individual categories, calculated based on our total sample of 202 ETGs.}
	\label{tab:Availability_of_merger_structures}
	\begin{tabular}{||l|c|c||} 
		\hline
		\textbf{Merger status} & \textbf{No of ETGs} & \textbf{Percentage} \\
		\hline
		\hline
        Relaxed (R) & 31 & 15\% \\
        \vspace{-0.3em}\\
		Minor merger (m) & 115 & 57\% \\
        \ \ \ \ $\bullet$ Mixed minor merger & 5 &  \\
        \ \ \ \ \ \ (m \& m) &  &  \\
        \vspace{-0.5em}\\
		Major merger (M) & 54 & 27\% \\
		\vspace{-0.3em}\\
        Mixed merger (m \& M)  & 2 & 1\% \\
		\hline
	\end{tabular}
\end{table}
In Section \ref{structural characteristics}, we described the various underlying apparent merger remnant structures seen in our sample of early-type galaxy images after removing the smooth stellar light distribution.
From Table~\ref{tab:Major_minor_structure_availability}, it is evident that shells are the dominant form of such structures found in early-type galaxies. Though most of these seem to result from minor mergers, a significant fraction is from major mergers. On the other hand, the rarest such structures seem to be the presence of embedded disk fragments. Boxy isophotes are also quite rare. 

The merger remnant structures can be arranged in ascending order, in terms of frequency of occurrence, as: disk fragments near the core (2\%), collisional ring structures (5\%), tidal tails (5\%), isophotal twisting (6\%), X-structure (7\%), disturbed galaxy (9\%), tidal fan (9\%), dust lanes (11\%), boxy \& disky isophotes (15\%), stellar streams \& plumes (17\%) and shell structures (54\%). 
We find that the number of galaxies with apparent minor merger remnant features is 57\% of the sample, and those with only major merger remnant features consist of  27\% of the sample (Table~\ref{tab:Availability_of_merger_structures}). Interestingly, two galaxies in our sample have both remnants characteristic of major and minor merger signatures. Several numerical simulations have suggested that frequent minor mergers can significantly alter the morphology and characteristics of a galaxy \citep[e.g.,][]{Bournaud2007,Karademir2019}. We have therefore indicated here the number of ETGs, found in our sample, that have indications of  consecutive minor mergers (i.e., presence of multiple signatures of individual minor merger scenarios) in Table~\ref{tab:Availability_of_merger_structures} (5 ETGs). These are intriguing cases, which we plan to follow up in future.

Out of 202 ETGs, only 31 galaxies (i.e. 15\%) were found to be in a relaxed state, with no evidence of underlying structures, a percentage consistent with the recent study by \citet{Tal2009,Kado-Fong2018,Jackson2021}. It indicates that ETGs are prone to having merger debris embedded in them, and one would need deep data sets to unravel them. It also supports the role of hierarchical merging and close interaction between galaxies in the evolution of early-type galaxies.

In order to assess the effectiveness of the image processing algorithms used in Section \ref{iat}, we attempted to identify these merger-related features, using the original images. We found that:
\begin{enumerate}
    \item Without the image processing techniques (i.e., \textit{model subtraction} and \textit{unsharp masking}), the proper identification of merger debris, in only 49.5\% of the galaxies (100 ETGs) in our sample, would have been possible by careful inspection of the original images  without any further processing. Many of these galaxies have the relevant structures on the outskirts of the galaxy and are thus easier to identify.  Features such as inner shells, central dust lanes, embedded disks and X-structures, which appear in the inner parts of the galaxies, are in general harder to identify. We found $\sim$20 ETGs having such inner structures that were completely not identifiable in the original images. 
    \item In an additional 11 galaxies (5.4\%), merger-related structures can be detected from the original unprocessed images, but they would have been misclassified due to the low contrast and confusion with diffuse radiation in the inner parts of the galaxies.
    \item For 71 galaxies (35.1\%), structures associated with mergers would not have been detected at all in the original images if the image processing were not performed.
    \item In 31 galaxies (15.3\%), no structures associated with mergers were detected even with the image processing techniques used in this paper (see Table~\ref{tab:Availability_of_merger_structures}).

\end{enumerate}
\noindent Thus, the number of galaxies with merger-related structures would have been seriously underestimated without the processing techniques used in this paper, leading to significant misinterpretation of the processes of transformation involved in the evolution of ETGs in the local universe.

\section{Classification of merger remnants} \label{Classification of merger remnants}
Considering the eleven types of merger structures recovered in our study, we have systematically tried to regroup them into categories related to their purported origin. It is similar to other attempts in classifying ETGs in sequences in their post-merger phase   \citep[e.g][]{Kaviraj2010,Cappellari2011, Duc2015}, which have not, in particular, focused on merger remnants. In this section, we intend to find a classification scheme for these merger structures based on the discussion carried out in Section \ref{structural characteristics}. 

We start from the classification proposed by \citet{Duc2015} and divide the ETGs according to:
\begin{itemize}
  \item Relaxed early-type (R)
  \item Recent minor merger (m)
  \item Recent major merger (M)
  \item Ongoing interaction (I)
\end{itemize}
The `I-type' here represents the galaxies involved in an ongoing interaction with a nearby companion, a characteristic not explicitly considered in this work. 
We define a parameter `$r$' that represents the ratio of the mass of the companion galaxy to the primary galaxy in an interaction event. The primary galaxy is defined as that of the higher mass, so the value of $r$ is always  $r\leq1$. We define minor mergers as those with  $r<0.7$, and major mergers with $0.7 \leq r \leq 1$ \citep[similar to][]{Mancillas2019,Wang2020}. Following the discussion in Section \ref{structural characteristics} and the references therein, we further subdivide the minor merger structures into three categories as 
\begin{itemize}
  \item Minor merger with a big mass ratio [\textbf{b}: $0.1 \leq r < 0.7$]
  \item Minor merger with an intermediate-mass ratio [\textbf{i}: $0.01 \leq r       <0.1$]
  \item Minor merger with a small mass ratio [\textbf{s}:  $r < 0.01$]
\end{itemize}
\begin{figure*}
\centering
	\includegraphics[width=1.3\columnwidth, height=12cm]{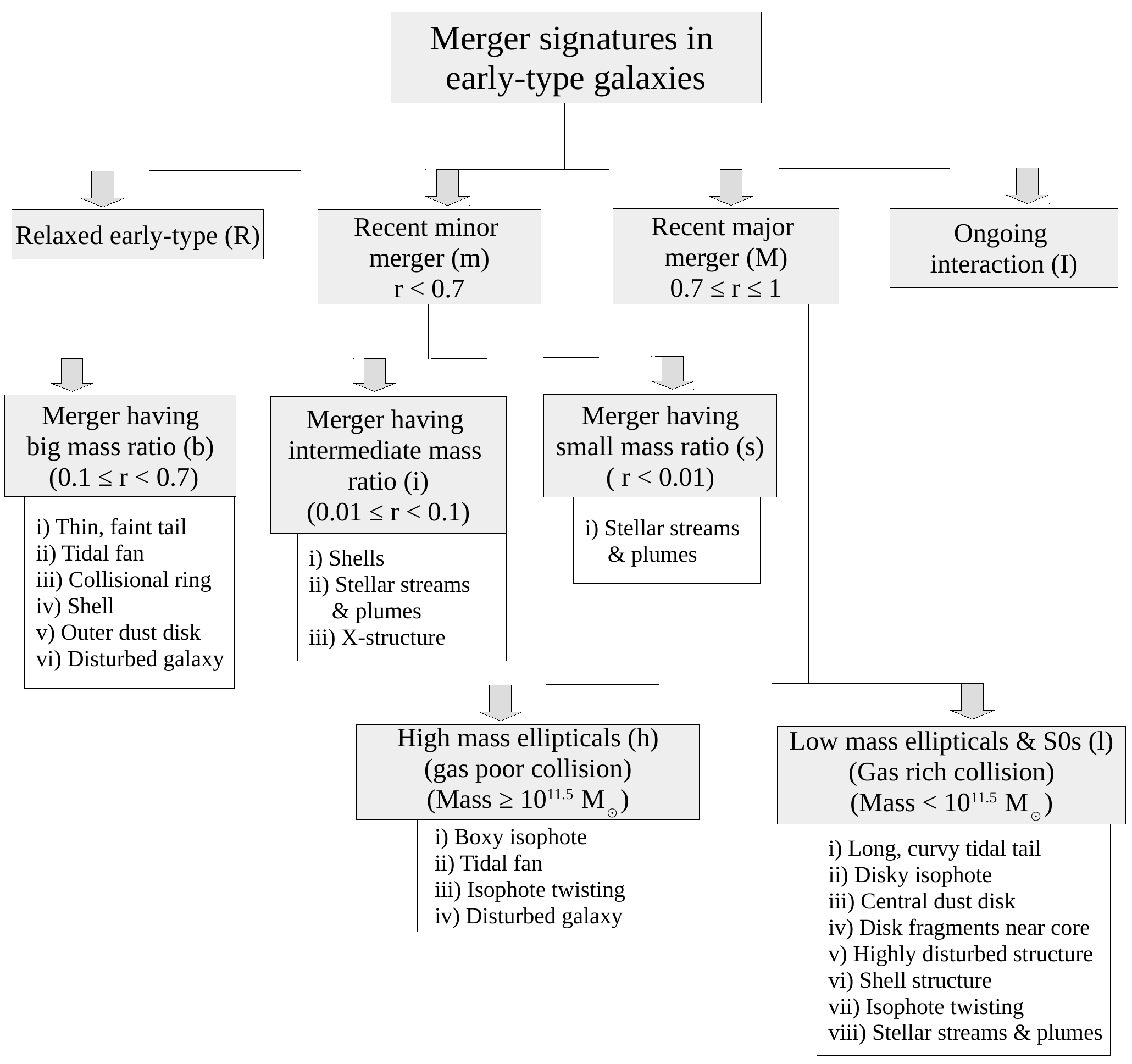}
    \caption{The classification scheme proposed in this study for the merger structures identified in this work. It is primarily based on the parameter `$r$', which is the ratio of the masses of the companion galaxy to the primary galaxy. The maximum value of $r$ is 1 for the major merger of similar masses. The structures associated with each class are also labelled in this table. Due to the multiple formation scenarios, some structures take multiple places in this classification scheme. See Section \ref{Classification of merger remnants} for more details.}
    \label{fig:table_of_classification}
\end{figure*}

The merger remnants in these categories are highlighted in the morphological diagram (Fig.~\ref{fig:table_of_classification}), where the major merger  structures are classified based on the mass of the resultant post-merger galaxy \cite[e.g.][]{Emsellem2011,Goullaud2018,Li2018} as: 
\begin{itemize}
  \item High mass ellipticals [\textbf{h}: mass $\geq\ 10^{11.5}M_{\odot}$]
  \item Low mass ellipticals and S0s [\textbf{l}: mass $< 10^{11.5}M_{\odot}$].
\end{itemize}
Since in major merger systems, the ratio `$r$' lies within a narrow range, i.e. $0.7\leq r \leq 1$,
it is challenging to classify remnants formed in such collisions based on $r$.
However, since massive ellipticals are thought to be the outcome of a series of gas-poor major mergers, whereas low mass ETGs primarily result from gas-rich disk collisions \citep{Naab2006, Loubser2022}, it is expected that the merger features in them can be classified into different categories.

In Fig.~\ref{fig:table_of_classification}, we summarize this classification scheme and list the merger structures associated with each category. However, since some merger structures can have multiple origins (see discussion in Section \ref{structural characteristics} and references therein), we have put them in multiple places in the tree. 
We assign merger types to ETGs based on the merger structures recovered in the galaxy by the methods used in this section. Based on this, we have prepared a `Merger Catalogue of Early-type Galaxies', shown in Table~\ref{tab:structure} (listing the first 20 galaxies; the complete list is available online). 
In the Table, we use the labels defined in Fig.~\ref{fig:table_of_classification}. 
For example, if one galaxy reveals an underlying structure corresponding to a  collisional ring, we attach the label `mb' to it in Column~4. Similarly, suppose a shell appears in a galaxy, in that case, we will label it with `mb' or `mi' (`mb/mi') or `Ml', depending on the structure's properties as discussed in Section \ref{structural characteristics}. This exercise can further help us understand in detail the history of hierarchical growth, and hence the evolutionary history, of an early-type galaxy by analysing the topology of the signatures of merger.
\begin{table*}
	\centering
	\caption{The first 20 Early-type Galaxies of our S82 sample. The entire catalogue is available in the electronic version of this paper.}
	\begin{center}
	\textbf{Merger Catalogue of Early-type Galaxies}
	\end{center}
	\label{tab:structure}
	\begin{tabular}{lcccl} 
		\hline
		\textbf{SDSS name} & \textbf{R.A. (deg.)} & \textbf{Dec. (deg.)} & \textbf{Structure} & \textbf{Comments}\\
		&\textbf{(J2000.0)}&\textbf{(J2000.0)}&\textbf{type}&\\
		\hline
		SDSS J204703.30+002612.4 & 311.7637795 & 0.4367973 & Ml & Presence of inner shell, central dust lane\\
		SDSS J204720.36+002902.6 & 311.8348699 & 0.4840561 & mb/mi & Presence of sharp inner shells\\
		SDSS J204638.09+002021.6 & 311.6587528 & 0.3393609 & mb & Presence of inner shell (NW), disturbed galaxy\\
		SDSS J204724.30+001802.9 & 311.8512636 & 0.3008238 & mi & Stellar plumes\\
		SDSS J204955.71+005358.5 & 312.4821376 & 0.8995915 & mb, Ml & Tidal fan (NW), central disk\\
		SDSS J205424.12-004831.5 & 313.6004983 & -0.8087651 & mb/mi & Faint inner shell (NE)\\
		SDSS J205838.24+005445.2 & 314.6593608 & 0.9125844 & mb & Presence of faint inner shell (E, SW), tidal fan (N)\\
		SDSS J212258.68+010136.1 & 320.7445214 & 1.0267028 & Ml & Central disky feature\\
		SDSS J212414.67+011336.2 & 321.0611412 & 1.2267543 & mi & Presence of faint inner shell (NW)\\
		SDSS J213500.39-003041.1 & 323.7516252 & -0.5114354 & mi & Sharp outer shell, stellar plumes, X-structure\\
		SDSS J213645.79+011456.4 & 324.1908000 & 1.2489987 & mi & Presence of faint shell (S)\\
		SDSS J214351.30-001917.0 & 325.9637928 & -0.3214058 & mb/mi & Presence of sharp inner shells\\
		SDSS J220455.93-003551.4 &	331.2330659 &	-0.5976184 & 	Ml 	&	Central disk, stellar plumes\\
		SDSS J220522.43-002944.6 &	331.3434586 &	-0.4957338 &	mi 		&	Presence of inner shell (W)\\
		SDSS J220401.82+003504.6 &	331.0076068 &	0.5846048 	&	Ml 	&		Disky central structure\\
		SDSS J220425.31+004255.4 &	331.1054674 &	0.7154213 	&	Ml 		&	Stellar plumes, central disk, isophotal twisting\\
		SDSS J220620.52+010715.7 &	331.5854966 &	1.1210486 	&	mi 	&		X-structure, multiple shell, stellar plumes\\
		SDSS J220638.46+011036.7 &	331.6602578 &	1.1769039 	&	Ml 	&	Faint shell (S), isophotal twisting, central disk\\
		SDSS J222614.63+004004.0 &	336.5609764 &	0.6677861 &		mb, mi 	&		Sharp inner and outer shells, Stellar stream, tidal fan\\
		SDSS J222904.69-011105.6 &	337.2695614 &	-1.1849097 &	mb 	&		Tidal fan (N), outer and inner shell (NE, SW)\\
		\hline
	\end{tabular}
 \\
	\raggedright Note:  Column (1) SDSS identifier, Column (2) \& (3) RA and Dec. in degree, Column (4) type of structure (as proposed in the classification scheme discussed in Section \ref{Classification of merger remnants}), Column (5) details of the merger-related structures associated with each galaxy. The directions N, S, E, W represent North, South, East and West, respectively, where north is up, and east is to the left.\\
\end{table*}
\section{Merger fraction of ETGs in the local universe} \label{Finding merger fraction of ETGs in the local universe}

In order to assess the abundance of evidence of merger-related substructure in early-type galaxies and to relate it to the rate of major and minor galaxy interactions in the local universe, one needs a firm estimate of observed merger fractions. In this section, we compare the merger fraction of galaxies estimated in the local universe from observational catalogues or simulations with the values estimated in this work. 

To evaluate the merger fraction from the merger rate of galaxies, we first need to know the typical lifetime of a structure associated with a major or minor merger. Following Section \ref{structural characteristics}, a minor merger can result in tidal tails or  X-structures, which are generally found to have a lifetime of 1--2 Gyr \citep{Mihos1995, Hibbard1995,Feldmann2008}, whereas such mergers can also result in shells, fans or collisional ring structures, which can typically have shorter time-scales of existence, around $10^8$--$10^9$ years \citep{Theys1976, Quinn1984, Combes1995, Madore2009,Barway2020,Martinez-Delgado2022}. The polar ring structures can have a substantially longer lifetime of $\sim$1.5 Gyr 
\citep[e.g.][]{Bekki1998}, compared to the collisional ring structures mentioned above.  
Among structures of major merger origin, for example, stellar streams and plumes associated with isolated field galaxies typically have 
a lifetime of a few billion years \citep[e.g.][]{Paudel2013}. Similarly, shell structures from major mergers have a characteristic lifetime of 2-3 Gyr \citep[e.g.][]{Mancillas2019}. Another common major merger remnant, disks embedded in the core of early-type galaxies, usually lasts 2-4 Gyr after the merger \citep{Naab2006,Graham2012}. The detection of isophotal twisting, characteristic of major mergers, or the presence of prominent central dust lanes, indicates that the major merger, responsible for the feature, occurred roughly 1--2 Gyr ago \citep{Tran2001, Sanders2015}. Based on this knowledge, we can evaluate the average lifetime of merger-related structures for major ($\rm{\langle T\rangle}_{major}$) and minor ($\rm {\langle T\rangle}_{minor}$) mergers as:
\begin{align}
    \rm{\langle T\rangle}_{major} =\frac{\sum_{i} n_{major}^i T_{major}^i}{\sum_{i} n_{major}^i}\\ 
    \rm{\langle T\rangle}_{minor} =\frac{\sum_{j} n_{minor}^j T_{minor}^j}{\sum_{j} n_{minor}^j}
\end{align}
where i, j represent the major and minor merger structure type, ${\rm n_{major}^i}$, ${\rm n_{minor}^j}$ represent the frequency of particular structures as found in our work (see Table \ref{tab:Major_minor_structure_availability}), and ${\rm T_{minor}^i}$, ${\rm T_{major}^j}$ represent the stability timescale of the respective feature in Gyr, as discussed above.

From these equations, we find that the structures resulting from minor mergers on an average exist for $\sim$1~Gyr ($\rm{\langle T\rangle}_{minor}$). In contrast, structures associated with major mergers typically exist for $\sim$2.3~Gyr ($\rm{\langle T\rangle}_{major}$). We note that the  
associated structures might have evolved and become too faint to be observationally detected for mergers occurring before the most recent one. However, some would be enhanced and distorted due to the dynamical effects of the latest merger or to interaction with newer merger structures  \citep{Casteels2013,Mancillas2019,Wang2020}.

Estimates of the rate of major mergers at zero redshift vary between 0.03 to 0.08 per Gyr \citep[e.g.][]{Mateus2008,Hopkins2010}, depending on the mass of the galaxy (i.e. typically between $10^{10}\! -\! 10^{12}\, M_{\odot}$). It corresponds to, in our terms, the value of the major merger fraction (i.e. the fraction of galaxies in which structures resulting from major interactions are seen), estimated as (merger rate per Gyr) * (average major merger structure lifetime in Gyr, i.e. $\rm{\langle T\rangle}_{major}$). It results in an estimated value 
of $7\!- \! 19$\% in the local universe. 
Similar results from directly estimated merger fractions in the local universe have been found in other surveys \citep[e.g.]{Patton2002,Bertone2009,Ventou2019}. In this regard, from our work highlighted in Section \ref{structural characteristics}, we showed an observed value of major merger fraction as 27\% (noted in Table \ref{tab:Availability_of_merger_structures}), which is slightly higher than the value analytically estimated here. 
For minor mergers with mass ratios $r\! \geq \! 0.01$, estimates can be found of merger rates at $z \approx 0$ varying from 0.3-0.5 per Gyr for intermediate mass galaxies, typically between $10^{10} \!-\! 10^{11.5} M_{\odot}$
\citep{Jian2012,Rodriguez-Gomez2015}.
The incidence of corresponding structures (or merger fraction) is quoted in the range of $30 \!-\! 50$\% (e.g., 0.3/Gyr * 1.0 Gyr, 
$\rm{\langle T\rangle}_{minor}$) * 100\%), which is also slightly lower than our estimate of 57\% (see Table \ref{tab:Availability_of_merger_structures}).

The slightly higher values of merger fraction obtained in this work, compared to those obtained analytically from the literature, is not unexpected. We have considered only ETGs in our sample, which are the end-products of a long hierarchical process of mass assembly \citep{Naab2007,Naab_IAUS2013,Mancillas2019,Wang2020}. Furthermore, we have analysed deeper optical images compared to most other surveys, and our identification of merger structures involves algorithms that are nominally better at finding such low-luminosity underlying structures than those found by careful visual inspection of original images, as is typically used in many other studies. 
\section{Comparison with theoretical predictions}\label{Synergy with theoretical predictions}
Theoretical perspectives on galaxy interactions are critical in understanding galaxy formation and evolution. In particular, numerical simulations help to understand the changing  morphologies of the emerging features in closely interacting and  post-merger galaxies. In the literature, such investigations have demonstrated that equal-mass mergers are important occurrences that lead to significant changes in the measurable  characteristics of galaxies. Minor mergers have been predicted to prevail over massive mergers in the evolution of galaxies because they are more frequent \citep{Bournaud2005,Bournaud2007,Karademir2019}. 

Dedicated numerical investigations were carried out to identify and characterize the structures emerging from mergers. One of the significant results from these was that the radial infall of satellites with low angular momentum leads to the formation of shells. In contrast, the infall of satellites with high angular momentum produces streams \citep{Johnston2008,Pop2018,Karademir2019}. If there is any confusion between a tail and a stream, their dimensions can be used to distinguish between the two, with streams being thinner, less intense, and having more complex morphology than tails \citep{Hibbard1995,Mancillas2019}. 
Numerical simulations involving these phenomena have led to a better understanding of the formation of ETGs, demonstrating how mergers of galaxies with similar masses can lead to post-merger systems with significantly disrupted stellar distributions over a specific lifetime.  As these features begin to relax, the gas and dust settles near the centre, resulting in disk-like characteristics at the core \citep{Ji2014,Wang2020}. Such investigations were also carried out to explain the potential causes of various complex merging features, such as the collisional ring, the X-structure, and isophotal twisting \citep{Theys1977,Hernquist1989,Bekki1998,Renaud2018,Lagos2022}. 

We have adopted the conclusions of such simulations, giving rise to certain observed structures in our statistically motivated sample of 202 ETGs. It has led us to recover 11 distinct types of merger debris located in and around these galaxies (see Section \ref{structural characteristics}). 

As previously indicated, numerical simulations have also significantly explored a wide variety of values for the mass ratio, helping to understand how mergers involving  different combinations of masses could be responsible for the genesis of the same fine features. As an illustration, prominent and symmetric tidal structures are likely to form in equal mass mergers, while asymmetric tidal structures are more common in mergers of unequal masses \citep{Toomre1972,Hibbard1995,Combes1995,Lotz2008}. Whenever such insights were available, they were adopted and referenced in Section \ref{structural characteristics} while discussing that specific structure.

The effects of projection, and of the limited range in surface brightness, on the detection of features also need to be highlighted here. While the prominence of some merger structures can vary with projection, they are unlikely to disappear, as demonstrated in several numerical studies \citep{Lotz2008,Ji2014,Mancillas2019,Karademir2019}. Structures such as shells are more susceptible to projection than tails or streams, but their detectability, in  morphological terms, remains largely unaffected \citep{Pop2017}. The surface brightness limit is the more important issue concerning identifying features. For several merging scenarios, simulations have shown that fewer instances of  merger debris around ETGs are detected in shallower images \citep{Ji2014,Mancillas2019}. Although the surface brightness limit used in this study, and the image reduction procedures, are seen to efficiently detect merger structures in our sample of ETGs, one can expect a slightly higher number of merger debris around these galaxies, particularly those formed by minor mergers such as shells or streams, if deeper images are used. For instance, \citet{Mancillas2019} have shown that for a surface brightness cut of $33$  mag arcsec$^{-2}$, rather than $29$  mag arcsec$^{-2}$,  $\sim$2 times more tidal streams and shells are expected around such galaxies. So, the numbers highlighted in Table \ref{tab:Major_minor_structure_availability} \& \ref{tab:Availability_of_merger_structures} may be modified when using deeper images from the future surveys using larger telescopes \citep[e.g.,][]{VanDokkum2014,valls2017}. However, the overall conclusions of our work will only be strengthened in any such cases, namely that most (if not all) of the ETGs in the local universe are actively evolving by merging with nearby (mostly minor) galaxies, resulting in various signatures of recent mergers around them.

\section{Summary} \label{Conclusions}
We searched for morphological signatures associated with major and minor mergers in deep images of early-type galaxies (ETGs) in the local universe ($z \leq 0.05$) from the SDSS of the {\it Stripe82} survey. We used the morphological classifications of {\it Galaxy Zoo 2} to define our initial sample of ETGs. However, during the analysis of the images, we identified significant misclassification (the Galaxy zoo classification was based on visual inspection of original SDSS images), resulting in an eventual sample size of 202 ETGs. 
The proximity of these ETGs means that their large enough angular size allows the large-scale smooth component of the light distribution can be removed to extract the underlying low-luminous merger structures. It is done using two different methods: unsharp masking on the 2D images, and subtracting an isophotal model representing the smooth luminosity. 
The key results obtained in our analysis of this sample are summarized below: 
\begin{enumerate}
\item From deep images of a  well-defined sample of early-type galaxies, we have identified  11 distinct kinds of structures associated with remnants of minor or major mergers. In terms of frequency of occurrence, in ascending order, they are: disk fragments near the core (2\%), collisional ring structures (5\%), tidal tails (5\%), isophotal twisting (6\%), X-shaped structures (7\%), disturbed galaxy (9\%), tidal fan (9\%), dust lanes (11\%), boxy \& disky isophotes (15\%), stellar streams \& plumes (17\%) and shell structures (54\%). Some of these relics are the result of both major and minor mergers, and as a result, the features resulting from different origins have slight differences in appearance. In this study, we have elaborately discussed such differences with examples from our sample. Overall, 85\% ETGs in our sample are found to have underlying merger morphologies, possessing either major (27\%) or minor (57\%) merger features. 1\% galaxies show the presence of individual major and minor merger signatures in them, indicating a possible recent mixed merger history.

To emphasize the significance of image reduction techniques in identifying merger structures, we have attempted to identify these features using the original images. While in 49.5\% of the galaxies in our sample, the merger-related features would have been detected by careful inspection of the original deep images, without the image processing techniques used in this paper, they would not have been detected, or misclassified if detected, in another 35.5\% of the galaxies. A majority of these structures remain well-hidden in the smooth dominant distribution of these galaxies, and a few structures cannot be properly identified due to the low contrast with their surroundings. Thus, in the absence of these image processing techniques, we would be significantly misled in the interpretation of the predominant processes driving the evolution of  ETGs in the local universe.

\item We find that shell structures are the dominant merger-related structures associated with early-type galaxies, accounting for 54\% of the galaxies in the S82 sample. Among these, the dominant source seems to be the
minor mergers; however, shells originating from major mergers are also significant too (24\% of the shell host ETGs). We conclude here that the frequency of occurrence of shell structures around ETGs is higher than hitherto found. It shows the importance of the requirement for deeper data sets in addition to the efficient removal of the smooth light distribution of the galaxy to properly uncover the inner and outer shells. 

The boxy morphology of the stellar distribution (1 out of 202 ETGs) and the presence of disk fragments near the core of the galaxy are the most uncommon merger structures seen in our sample. The former indicates that mergers between massive ellipticals are rare in the local universe, whereas the latter indicates that more simulations and theoretical investigation are needed to understand the merger process leading to such rare and peculiar remnants. Our study also shows the need for further understanding of merger processes that create isophotal twisting in ETGs, since these are rarer than expected. We should also mention that galaxies with isophotal twisting are more likely to host multiple shells ($>3$) in them, contrary to what has been reported earlier.
\item The main aim of this work is not just to estimate the frequency of occurrence of merger remnants within ETGs, but also to develop a classification scheme for these fine structures that may be used to infer their likely history and genesis. First, we divide the ETGs into four basic classes: Relaxed ETGs (R), ETGs with recent minor mergers (m), ETGs with recent major mergers (M) and the ETGs in Interacting systems (I). Here, we use the parameter `$r$' (ratio of the masses of the pre-merger galaxies) to differentiate between minor and major mergers. We further divide the minor merger structures into three subclasses based on this parameter. On the other hand, we divide the major merger structures into two subclasses based on the mass of the resulting merger product (it is difficult to subdivide the major merger structures based on `$r$', since a narrow range of $r$ is available). The merger remnants with specific morphologies, indicative of distinct past interaction histories, are then listed in their respective classes.
\item Finally, the major and minor merger fractions of galaxies (i.e., the number of merged galactic systems) obtained by us in the local universe are significantly higher than results obtained in previous studies. It can partly be attributed to the fact that our sample consists only of ETGs (which remain at the end of the mass assembly process) and the use of the image processing techniques (which helped in better detecting the buried structures) that we use of the deepest images that are available in the region of our sample.
\\
\\
This study concludes that a high fraction of galaxies in the local Universe, with early-type morphologies, have had recent mergers, and the signatures of such mergers can be found in and around these galaxies. We have discovered shells and streams to be the most dominant merger features of ETGs, which supports the findings of several studies of numerical simulation. However, such studies also predict the detection of a greater number of merger debris (e.g., shells and streams) around these galaxies if deeper images are used than those used here \citep{Mancillas2019}. In this regard, the merger rate of ETGs and the stability timescale of these detected structures, found in the present study, indicate the possibility of detecting a significant number of merger debris around these galaxies, providing support to the initial idea. As a result, we can anticipate that most (if not all) ETGs in the nearby Universe are constantly evolving through mergers, mainly minor mergers. The new deep surveys from large-aperture observing facilities will achieve lower surface brightness limits than those used in this work, which will further help in quantifying the incidence of these structures.

\end{enumerate}

\section*{Acknowledgements}
The work presented here has used the Inter-University Centre for Astronomy and Astrophysics (IUCAA) high-performance computing facilities in Pune, India, and the Indian Institute of Technology, Indore, India. SB and GG thank IUCAA for their hospitality and the use of facilities without which this work could not have been done. GG is supported by a Prime Minister's Research Fellowship.

\section*{Data Availability}
The archival $r$-band data used in this work are available under the `IAC Stripe 82 Legacy Project': \url{http://research.iac.es/proyecto/stripe82/}. The 55 bright elliptical galaxies used as the control sample are available here (the OBEY survey): \url{http://www.astro.yale.edu/obey/index.html}. We publish the `Merger Catalogue of Early-type Galaxies' as a product of this work, which is available online with this article.



\bibliographystyle{mnras}
\bibliography{ref} 

%

%





\bsp	
\label{lastpage}
\end{document}